%% file: main.tex
\documentclass[lettersize,journal]{IEEEtran}

\usepackage{amsmath,amsfonts,amssymb}
\usepackage{algorithmic}

\usepackage{graphicx}
\usepackage{array}
\usepackage{multirow}
\usepackage{booktabs}
\usepackage{makecell}
\usepackage{diagbox}
\usepackage{tabularx}
\newcolumntype{Y}{>{\centering\arraybackslash}X}

\usepackage[caption=false,font=normalsize,labelfont=sf,textfont=sf]{subfig}

\usepackage{textcomp}
\usepackage{verbatim}
\usepackage{lipsum}
\usepackage{changepage}
\usepackage{stfloats}
\usepackage{balance}
\usepackage{bm}
\usepackage{url}
\usepackage[compress]{cite}
\usepackage[colorlinks,urlcolor=blue,citecolor=blue,linkcolor=blue]{hyperref}

\def\BibTeX{{\rm B\kern-.05em{\sc i\kern-.025em b}\kern-.08em
    T\kern-.1667em\lower.7ex\hbox{E}\kern-.125emX}}

\newcommand{\mt}[1]{\text{#1}}

\begin{document}


\title{\vspace{-0.5em} Neural Two-Stage Stochastic Volt-VAR Optimization for Three-Phase Unbalanced Distribution Systems with Network Reconfiguration}
\author{Zhentong Shao,~\IEEEmembership{Member,~IEEE}, Jingtao Qin,~\IEEEmembership{Student Member,~IEEE}, Nanpeng Yu,~\IEEEmembership{Senior Member,~IEEE}
\vspace{-2.5em}
}

\maketitle

\begin{abstract}
The increasing integration of intermittent distributed energy resources (DERs) has introduced significant variability in distribution networks, posing challenges to voltage regulation and reactive power management. 
This paper presents a novel neural two-stage stochastic Volt-VAR optimization (2S-VVO) method for three-phase unbalanced distribution systems considering network reconfiguration under uncertainty. To address the computational intractability associated with solving large-scale scenario-based 2S-VVO problems, a learning-based acceleration strategy is introduced, wherein the second-stage recourse model is approximated by a neural network. This neural approximation is embedded into the optimization model as a mixed-integer linear program (MILP), enabling effective enforcement of operational constraints related to the first-stage decisions. Numerical simulations on a 123-bus unbalanced distribution system demonstrate that the proposed approach achieves over 50 times speedup compared to conventional solvers and decomposition methods, while maintaining a typical optimality gap below 0.30\%. These results underscore the method’s efficacy and scalability in addressing large-scale stochastic VVO problems under practical operating conditions.
\end{abstract}


\vspace{-1em}
\section{Introduction}

\subsection{Research Motivation}

The growing integration of distributed energy resources (DERs), especially intermittent sources such as photovoltaic (PV) systems, has introduced significant operational uncertainties in distribution networks. These uncertainties pose major challenges for maintaining stable voltage profiles and managing reactive power efficiently. To address these issues, two-stage stochastic volt-var optimization (2S-VVO) has emerged as an effective tool for decision-making under uncertainty \cite{ding2015two}. 
The basic 2S-VVO problem typically considers the decisions of on-load tap changers (OLTCs) and capacitor banks in a balanced three-phase system. However, for practical applications, it is more relevant to address the 2S-VVO problem within unbalanced three-phase distribution systems incorporating network reconfiguration. This extension significantly increases the complexity of the problem, particularly when stochastic programming is employed to account for a large number of scenarios \cite{frau2012stochastic}. The combinatorial nature of network reconfiguration and the nonlinearity of three-phase power flow further exacerbate the computational burden. Consequently, solving the 2S-VVO problem using conventional approaches becomes impractical for large-scale or real-time implementations.
 

In this paper, we first establish a comprehensive 2S-VVO model, where the first stage optimizes the decisions of OLTCs and remotely controlled switches, and the second stage, formulated as a recourse problem, determines the outputs of DERs. In particular, voltage violation duration constraints are integrated into the second-stage recourse problem, imposing more stringent and practical voltage stability requirements. As a result, the second-stage problem becomes a mixed-integer linear program (MILP), which further increases the computational complexity of solving the overall 2S-VVO problem. 

To alleviate the computational burden of the proposed 2S-VVO model, this paper introduces a novel neural two-stage stochastic optimization framework, in which the second-stage recourse problem is approximated by a neural network surrogate. The surrogate is reformulated as a small-scale MILP model, allowing efficient integration into the first-stage optimization. This enables explicit enforcement of hard constraints associated with first-stage decisions, representing a key advantage over conventional direct-mapping methods. Besides, the proposed neural optimization method also functions as a learning-based scenario reduction strategy, which avoids the linear growth of model size with respect to the number of stochastic scenarios. As a result, it effectively captures system uncertainty while maintaining tractability. Numerical experiments show that the method achieves over 50-fold speedup compared to conventional solvers, while maintaining a typical optimality gap of less than 0.30\%.

\vspace{-0.5em}
\subsection{Literature Review}
\subsubsection{Volt-Var Optimization}
The VVO is a critical mechanism in regulating distribution systems' voltage profiles and minimizing the curtailment of DERs. Conventional VVO methods primarily target single-phase balanced distribution systems, relying on the control of OLTCs and the reactive power outputs of compensators \cite{salih2015coordinated,hashemi2016efficient}. Recent studies have extended VVO frameworks to accommodate more realistic three-phase unbalanced distribution systems and incorporate additional active distribution network (ADN) management strategies, such as dynamic network reconfiguration \cite{wu2018distribution,kianmehr2019resilience}. Jointly optimizing network reconfiguration, OLTC operations, and DER coordination has been shown to significantly enhance voltage stability and operational efficiency, particularly in unbalanced practical distribution systems \cite{chen2017robust, qin2025physics}. However, such comprehensive formulations substantially enlarge the problem scale and computational burden, posing challenges for real-world implementation.

To mitigate the computational complexity of VVO models, researchers have proposed decomposition and distributed optimization methods, with the Alternating Direction Method of Multipliers (ADMM) being a representative and widely adopted approach. Specifically, a distributed three-block ADMM-based framework is proposed to solve the DER coordination problem under VVC constraints in distribution networks \cite{gebbran2022multiperiod}. A bi-level ADMM-based method is proposed to enhance convergence and reduce communication overhead in distributed Volt/Var optimization for ADNs \cite{ju2021bi}. An accelerated ADMM-based method is developed for fully distributed reactive power optimization, incorporating a quasi-Newton strategy to enhance convergence with minimal information exchange between areas \cite{xu2020accelerated}. Notably, decomposition and distributed optimization frameworks have been widely adopted to manage computational complexity, often emphasizing challenges such as communication latency and data privacy. In contrast, this paper introduces a fundamentally different strategy -- an acceleration approach based on a neural network-embedded optimization framework.

Another important aspect of VVO is the modeling of uncertainty, which can generally be classified into two main categories: robust optimization \cite{ding2015two, zhou2021three} and scenario-based stochastic optimization \cite{leng2023two, wang2022two}. Robust optimization produces reliable solutions under worst-case conditions but often results in overly conservative dispatch strategies, leading to unnecessary DER curtailment. Moreover, it significantly increases computational complexity, particularly in large-scale, three-phase unbalanced distribution systems. To mitigate this conservativeness and improve scalability, stochastic VVO approaches \cite{santos2016new, al2017probabilistic} are frequently adopted. These methods generate multiple scenarios to capture uncertainties, featuring simpler modeling structures that accommodate detailed device representations and scale more effectively to large networks. However, their computational complexity grows linearly with the number of scenarios \cite{lee2025column}. To address these drawbacks, two primary acceleration strategies have been developed: optimization-based decomposition methods \cite{zeng2013ccg, oliveira2014accelerating} and scenario reduction techniques \cite{zhang2023optimized}.

Considering the inherent complexity of three-phase unbalanced VVO problems -- particularly when network reconfiguration is involved -- this study focuses on accelerating optimization under a large number of scenarios. In contrast to traditional acceleration strategies, we propose a learning-based acceleration framework that leverages deep learning to extract and represent key uncertainty features. By embedding neural networks within the optimization process, we transform a large-scale, scenario-dependent VVO problem into a compact MILP formulation. This enables the inclusion of extensive uncertainty scenarios while maintaining computational efficiency and solution quality. The proposed framework provides a substantial advancement over conventional acceleration methods, combining scalability with improved optimality guarantees.

\subsubsection{Neural Network-Based Optimization}
The core technique of this paper involves embedding neural networks into optimization problems, a direction that has recently emerged as a powerful paradigm. Foundational work \cite{cheng2017maximum,tjeng2017evaluating} demonstrated that neural networks with Rectified Linear Unit (ReLU) activations can be reformulated as MILP models, enabling exact representations of their decision boundaries. This line of research was further extended to adversarial learning applications \cite{fischetti2018deep}, with subsequent refinements such as calibrated big-M constraints and improved treatment of inactive neurons enhancing tractability and solution strength \cite{grimstad2019relu,anderson2020strong}. The development of the Optimization and Machine Learning Toolkit (OMLT) \cite{ceccon2022omlt} has further advanced scalability, supporting complex architectures such as Graph Neural Networks (GNNs). Beyond methodological progress, this modeling paradigm has been successfully applied across diverse domains such as market bidding \cite{jalving2023beyond}, reinforcement learning \cite{shengren2023optimal}, and stochastic programming \cite{patel2022neur2sp}, underscoring its versatility and growing practical relevance.

\vspace{-1em}
\subsection{Contribution}

This paper proposes a 2S-VVO model considering network reconfiguration for three-phase unbalanced distribution systems. To tackle the computational challenges of solving large-scale 2S-VVO problems under numerous uncertainty scenarios, we propose a neural two-stage stochastic optimization framework to accelerate the solution process. The main contributions of this paper are summarized as follows:
\begin{enumerate}
    \item[1)] A two-stage stochastic VVO model is developed for three-phase unbalanced distribution systems with network reconfiguration. To enhance voltage compliance under uncertainty, we introduce a novel voltage violation duration constraint that explicitly limits the cumulative time of voltage magnitude deviations. This formulation moves beyond conventional snapshot-based limits, providing a more practical safeguard against persistent violations and improving system reliability under stochastic operating conditions.

    \item[2)] A neural two-stage stochastic VVO solution framework is developed to efficiently solve large-scale 2S-VVO problems with numerous uncertainty scenarios. In this framework, the second-stage recourse problem is approximated using a deep neural network, which is reformulated as an MILP and embedded into the first-stage model, enabling explicit enforcement of hard constraints associated with network reconfiguration decisions while ensuring computational scalability and maintaining near-optimal performance.

    \item[3)] The proposed framework achieves significant computational acceleration by decoupling model complexity from the number of scenarios. Numerical results show that the framework maintains high solution quality, with an optimality gap below 0.30\%, while reducing computation time by up to two orders of magnitude and achieving an average speedup of over 50$\times$.
\end{enumerate}

The remainder of the paper is organized as follows. Section II introduces the proposed 2S-VVO model. Section III presents the proposed neural two-stage stochastic optimization framework. Section IV discusses case studies and performance evaluation. Finally, Section V concludes the paper.

\section{Two-Stage Stochastic Volt-Var Optimization Model with Network Reconfiguration}

\subsection{Three-Phase VVO Problem with Network Reconfiguration}
This section presents a comprehensive VVO model for three-phase unbalanced systems with network reconfiguration. 

\subsubsection{Objective Function}
The objective is to minimize the weighted sum of the total active power curtailment of DERs and the operational frequency of switches and OLTCs.

\vspace{-4mm}
\begin{gather}
\begin{aligned}
    \text{Obj}. \ \underset{Pg,\gamma,\rho}{\min} &\sum_{t \in \mathcal{T}} \left( \sum_{i \in \mathcal{G}} \sum_{\varphi \in \psi_i} \omega^{\mt{der}} \left( \widehat{Pg}_{i,t}^\varphi - Pg_{i,t}^\varphi \right) \right. \\ 
    &\left. + \sum_{(i,j) \in \mathcal{E}^{\mt{sw}}} \omega^{\mt{sw}} \gamma_{ij,t} + \sum_{(i,j) \in \mathcal{E}^{\mt{oc}}} \omega^{\mt{oc}} \rho_{ij,t} \right),
    \label{eq:obj}
\end{aligned}
\end{gather}
\vspace{-2mm}

\noindent where $\mathcal{T}$ is the set of time periods. $\widehat{Pg}_{i,t}^\varphi$ and $Pg_{i,t}^\varphi$ represent the maximum available and dispatched active power from the DER at bus $i$ and phase $\varphi$ at time $t$, respectively. $\mathcal{G}$ is the set of buses equipped with DERs. $\psi_{i}$ denotes the set of phases at bus $i$. $\mathcal{E}^{\mt{sw}}$ and $\mathcal{E}^{\mt{oc}}$ denote the sets of branches equipped with switches and OLTCs, respectively. $\omega^{\mt{der}}$ denotes the cost coefficient associated with DER curtailment, while $\omega^{\mt{sw}}$ and $\omega^{\mt{oc}}$ represent the operational cost coefficients corresponding to switches and OLTCs, respectively. $\gamma_{ij,t}$ and $\rho_{ij,t}$ represent the switching and tap-changing actions for branch $(i,j)$ at time $t$. The VVO with network reconfiguration is subject to the following constraints.

\subsubsection{Nodal Power Balance Constraints}
\begin{gather}
    \left\{
        \begin{aligned}
            & P_{ij,t}^\varphi = \sum_{k:j \rightarrow k} P_{jk,t}^\varphi + \widehat{Pd}_{j,t}^\varphi - Pg_{j,t}^\varphi, \\
            & Q_{ij,t}^\varphi = \sum_{k:j \rightarrow k} Q_{jk,t}^\varphi + \widehat{Qd}_{j,t}^\varphi - Qg_{j,t}^\varphi,
            \\
            &  ~\forall \varphi \in \psi_j,\ \forall j \in \mathcal{N}^{+},\ \forall t \in \mathcal{T}.
        \end{aligned}
    \right.
    \label{eq:Balance}
\end{gather}
where $\mathcal{N}^{+}$ denotes the set of non-substation buses. $P_{ij,t}^{\varphi}$ and $Q_{ij,t}^{\varphi}$ denote the active and reactive power flow from bus $i$ to bus $j$ at time $t$, respectively. $\widehat{Pd}_{j,t}^{\varphi}$ and $\widehat{Qd}_{j,t}^{\varphi}$ denote the active and reactive power load at bus $j$ at time $t$, respectively. $Qg_{i,t}^\varphi$ is the reactive power output of DER at bus $i$ and time $t$.

\subsubsection{Branch Power Flow Constraints}
\begin{gather}
\left\{
\begin{aligned}
    &\bm{U}_{j,t} \leq \bm{U}_{i,t} + \bm{M}^{P}_{ij} \bm{P}_{ij,t} + \bm{M}^{Q}_{ij} \bm{Q}_{ij,t} + M (1 - \alpha_{ij,t}) \bm{1},
    \\
    &\bm{U}_{j,t} \geq \bm{U}_{i,t} + \bm{M}^{P}_{ij} \bm{P}_{ij,t} + \bm{M}^{Q}_{ij} \bm{Q}_{ij,t} - M (1 - \alpha_{ij,t}) \bm{1},
    \\
    & \forall (i,j) \in \mathcal{E} \setminus \mathcal{E}^{\mt{oc}},\ \forall t \in \mathcal{T}.
\end{aligned}
\label{eq:PowerFlow}
\right.
\end{gather}
where $\bm{U}_{j,t}$, $\bm{P}_{ij,t}$, and $\bm{Q}_{ij,t}$ denote the three-phase vectors corresponding to $U_{j,t}^{\varphi}$, $P_{ij,t}^{\varphi}$, and $Q_{ij,t}^{\varphi}$, respectively. $U_{j,t}^{\varphi}$ denotes the squared voltage magnitude. $\bm{M}^{P}_{ij}$ and $\bm{M}^{Q}_{ij}$ are defined in \eqref{eq:MPQCal}. The binary variable $\alpha_{ij,t}$ indicates the connection status of branch $(i,j)$ at time $t$. The set $\mathcal{E}$ comprises all branches in the network, while $\mathcal{E}^{oc}$ denotes the subset of branches equipped with OLTC. The scalar $M$ represents a sufficiently large constant, and $\bm{1}$ is a vector of ones. \eqref{eq:PowerFlow} is derived from the three-phase LinDistFlow model \cite{arnold2016optimal}.

\begin{gather}
\left\{
\begin{aligned}
    &\bm{M^{P}}_{ij} =
        \begin{bmatrix}
        -2r_{aa} & r_{ab} - \sqrt{3}x_{ab} & r_{ac} + \sqrt{3}x_{ac} \\
        r_{ba} + \sqrt{3}x_{ba} & -2r_{bb} & r_{bc} - \sqrt{3}x_{bc} \\
        r_{ca} - \sqrt{3}x_{ca} & r_{cb} + \sqrt{3}x_{cb} & -2r_{cc}
        \end{bmatrix}_{ij}
    \\
    &\bm{M^{Q}}_{ij} =
        \begin{bmatrix}
        -2x_{aa} & x_{ab} + \sqrt{3}r_{ab} & x_{ac} - \sqrt{3}r_{ac} \\
        x_{ba} - \sqrt{3}r_{ba} & -2x_{bb} & x_{bc} + \sqrt{3}r_{bc} \\
        x_{ca} + \sqrt{3}r_{ca} & x_{cb} - \sqrt{3}r_{cb} & -2x_{cc}
        \end{bmatrix}_{ij}
\end{aligned}
\label{eq:MPQCal}
\right.
\end{gather}
\noindent where $r_{aa}$ and $x_{aa}$ denote the self-resistance and self-reactance of phase $a$, while $r_{ab}$ and $x_{ab}$ represent the mutual resistance and mutual reactance between phases $a$ and $b$. The same notation applies correspondingly to the other symbols.


\subsubsection{Branch Power Flow with OLTC} To model the OLTC, a dummy bus $i'$ is introduced as illustrated in Fig. 2 of \cite{qin2023reconfigure}. Thus, the power flow model for branches with OLTC is given as below.

\vspace{-3mm}
\begin{gather}
\left\{
\begin{aligned}
    &\bm{U}_{j,t} \leq \bm{U}_{i',t} + \bm{M}^{P}_{ij} \bm{P}_{ij,t} + \bm{M}^{Q}_{ij} \bm{Q}_{ij,t} + M (1 - \alpha_{ij,t}) \bm{1},
    \\
    &\bm{U}_{j,t} \geq \bm{U}_{i',t} + \bm{M}^{P}_{ij} \bm{P}_{ij,t} + \bm{M}^{Q}_{ij} \bm{Q}_{ij,t} - M (1 - \alpha_{ij,t}) \bm{1},
    \\
    & \forall (i,j) \in \mathcal{E}^{\mt{oc}},\ \forall t \in \mathcal{T}.
\end{aligned}
\right.
\end{gather}
where $U_{i',t}^\varphi$ is defined using a Special Ordered Set of Type 1 (SOS1) formulation:
\begin{gather}
    U_{i',t}^\varphi = U_{i,t}^\varphi \cdot \sum_{m=1}^{M_{ij}} x_{ij,t}^m (\eta_{ij}^m)^{2},~~\sum_{m=1}^{M_{ij}} x_{ij,t}^m = 1.\label{eq:VR1}
\end{gather}
where $\eta_{ij}^{m}$ denotes the $m$-th tap ratio of the OLTC on branch $(i,j)$, $M_{ij}$ is the total number of available tap positions, and $x_{ij,t}^m$ is a binary variable corresponding to the selection of $m$-th tap setting at time $t$. The remaining nonlinear term $y_{ij,t}^{m,\varphi} = U_{i,t}^\varphi \cdot x_{ij,t}^m$ can be linearized via the big-M method in \eqref{eq:VR3}.
\begin{gather}
    \! \! 0 \leq y_{ij,t}^{m,\varphi} \leq M \cdot x_{ij,t}^m,~0 \leq U_{i,t}^\varphi - y_{ij,t}^{m,\varphi} \leq M \!\cdot\! (1 - x_{ij,t}^m) \label{eq:VR3}
\end{gather}

\subsubsection{Operational Restriction of OLTCs} The total number of OLTC tap operations is constrained to prevent excessive switching.
\begin{gather}
\left\{
\begin{aligned}
    &\sum_{m=1}^{M_{ij}} x_{ij,t}^m \cdot m = \tau_{ij,t}
    \\
    &|\tau_{ij,t} - \tau_{ij,t-1}| \leq \rho_{ij,t},~~\sum_{t \in \mathcal{T}} \rho_{ij,t} \leq \rho_{ij}^{\max},
\end{aligned}
\right.
\end{gather}
where $\tau_{ij,t}$ denotes the tap position of the OLTC on branch $(i,j)$ at time $t$, and $\rho_{ij,t}$ represents the number of tap changes at time $t$. The parameter $\rho_{ij}^{\max}$ specifies the maximum allowable daily operation count for the OLTC on branch $(i,j)$.

\subsubsection{DER Output Constraints}
\begin{gather}
\left\{
\begin{aligned}
    & 0 \leq Pg_{i,t}^\varphi \leq \widehat{Pg}_{i,t}^\varphi,\\
    & -Sg_{i}^\varphi \leq Pg_{i,t}^\varphi \leq Sg_{i}^\varphi,~-Sg_{i}^\varphi \leq Qg_{i,t}^\varphi \leq Sg_{i}^\varphi, \\
    & -\sqrt{2} Sg_{i}^\varphi \leq Pg_{i,t}^\varphi + Qg_{i,t}^\varphi \leq \sqrt{2} Sg_{i}^\varphi, \\
    & -\sqrt{2} Sg_{i}^\varphi \leq Pg_{i,t}^\varphi - Qg_{i,t}^\varphi \leq \sqrt{2} Sg_{i}^\varphi, \\
    &\forall \varphi \in \psi_i,\ \forall i \in \mathcal{G},\ \forall t \in \mathcal{T} \label{eq:DER_constr}
\end{aligned}
\right.
\end{gather}
where $Sg_{i}^\varphi$ denotes the apparent power capacity of the DER at bus $i$. The original quadratic capacity constraint is linearized using the circle constraint linearization method \cite{chen2015robust}. This linearization technique is also extended to effectively represent the thermal capacity limits of the distribution lines.

\subsubsection{Range B Voltage Limit Constraints}
\begin{gather}
\left\{
\begin{aligned}
    & U_{i,t}^\varphi = U_{\text{ref}}, ~ \varphi \in \psi_i,\ \forall i \in \mathcal{N}_{\mt{sub}},\ \forall t \in \mathcal{T}, \\
    & U_i^{\min} \leq U_{i,t}^\varphi \leq U_i^{\max}, ~ \varphi \in \psi_i,\ \forall i \in \mathcal{N}^{+},\ \forall t \in \mathcal{T}
\end{aligned}
\right.
\label{eq:voltage_limits}
\end{gather}
where $U_{\mt{ref}}$ is the reference squared voltage magnitude of the substation, and $\mathcal{N}_{\mt{sub}}$ denotes the set of substation buses. $U_i^{\min}$ and $U_i^{\max}$ represent the lower and upper bounds of the Range B voltage limits, i.e., $[0.917^{2}, 1.058^{2}]$ p.u., as specified in the ANSI C84.1-2016 standard~\cite{ANSI_C84_1_2016}.

Furthermore, to enhance voltage stability, the voltage magnitude imbalance at three-phase buses is constrained, as formulated in \eqref{eq:voltage_limits_3phase}, where $\mathcal{N}_{\mt{3-Phase}}$ is the set of buses with three-phase connections, and $\epsilon$ denotes the maximum allowable phase voltage deviation relative to the average.
\begin{gather}
\left\{
\begin{aligned}
    & -\epsilon \leq \frac{U_{i,t}^\varphi - U_{i,t}^{\mt{avg}}}{U_{i,t}^{\mt{avg}}} \leq \epsilon,~~
    U_{i,t}^{\mt{avg}} = \frac{1}{3} \sum_{\varphi \in \psi_i} U_{i,t}^\varphi, \\
    &\forall \varphi \in \psi_i,\ \forall i \in \mathcal{N}_{\mt{3-Phase}},\ \forall t \in \mathcal{T}
\end{aligned}
\right.
\label{eq:voltage_limits_3phase}
\end{gather}

\subsubsection{Range A Voltage Limit Constraints}
In addition to static voltage magnitude constraints, the dynamic voltage stability is also considered. Specifically, the cumulative duration of voltage violations and the maximum allowable length of consecutive violations are constrained as follows:
\begin{gather}
\left\{
\begin{aligned}
&\underline{U}_{i} - \delta_{i,t}^{\varphi} \leq U_{i,t}^{\varphi} \leq \overline{U}_{i} + \delta_{i,t}^{\varphi}, \\
&0 \leq \delta_{i,t}^{\varphi} \leq M \cdot z_{i,t}^{\varphi},\ \forall \varphi \in \psi_i,\ \forall i \in \mathcal{N}_d,\ \forall t \in \mathcal{T}
\\
&\sum_{t\in \mathcal{T}}  z_{i,t}^\varphi \leq d_1,\ \forall \varphi \in \psi_i,\ \forall i \in \mathcal{N}_d, 
\\
&\sum_{\tau = t}^{t + d_2}  z_{i,\tau}^\varphi \leq d_2,\ \forall \varphi \in \psi_i,\ \forall i \in \mathcal{N}_d,\ \forall t \in \mathcal{T}_d,
\end{aligned} 
\right.
\label{eq:duration_formulation}
\end{gather}
where $\overline{U}_i$ and $\underline{U}_i$ denote the Range A voltage limits as specified in the ANSI standard \cite{ANSI_C84_1_2016}, set to $[0.95^{2}, 1.05^{2}]$ p.u.. $\delta_{i,t}^\varphi$ is an auxiliary variable representing the deviation of the voltage magnitude at bus $i$ at time $t$ for phase $\varphi$. $z_{i,t}^\varphi$ is a binary variable indicating whether a voltage violation occurs. The parameter $d_1$ defines the maximum allowable total number of violation time steps over the entire scheduling horizon, while $d_2$ limits the maximum duration of consecutive violations. $\mathcal{N}_d$ and $\mathcal{T}_d$ denote the sets of monitored buses and monitored time steps, respectively.

\subsubsection{Branch Thermal Capacity Constraints}
\begin{gather}
\left\{
\begin{aligned}
    & -\alpha_{ij,t}S_{ij}^{\max} \leq P_{ij,t}^\varphi \leq \alpha_{ij,t}S_{ij}^{\max},  
    \\
    & -\alpha_{ij,t}S_{ij}^{\max} \leq Q_{ij,t}^\varphi \leq \alpha_{ij,t}S_{ij}^{\max},  
    \\
    & -\sqrt{2}\alpha_{ij,t}S_{ij}^{\max} \leq P_{ij,t}^\varphi + Q_{ij,t}^\varphi \leq \sqrt{2}\alpha_{ij,t}S_{ij}^{\max},  
    \\
    & -\sqrt{2}\alpha_{ij,t}S_{ij}^{\max} \leq P_{ij,t}^\varphi - Q_{ij,t}^\varphi \leq \sqrt{2}\alpha_{ij,t}S_{ij}^{\max},
\end{aligned} \label{eq:thermal_limit}
\right.
\end{gather}
where $S_{ij}^{\max}$ is the thermal capacity of branch $(i,j)$.

\subsubsection{Network Reconfiguration Constraints}
The network reconfiguration operations must adhere to three primary constraints: radiality, connectivity, and switch operation limits. First, the radiality constraint ensures that the network remains loop-free, as defined in \eqref{eq:radiality}.
\begin{gather}
    \sum_{(i,j) \in \mathcal{E}} \nolimits \alpha_{ij,t} = |\mathcal{N}| - 1,\quad \forall t \in \mathcal{T}
    \label{eq:radiality}
\end{gather}

Next, the connectivity constraint ensures that each bus (excluding substations) is connected to exactly one upstream bus. This is formulated in \eqref{eq:connectivity-1} and \eqref{eq:connectivity-2}.
\begin{gather}
\beta_{ij,t} + \beta_{ji,t} = \alpha_{ij,t},\quad \forall (i,j) \in \mathcal{E},\ \forall t \in \mathcal{T}
\label{eq:connectivity-1}
\\
\sum_{j:j \rightarrow i} \beta_{ji,t} = 
\begin{cases}
    1, & \forall i \in \mathcal{N}^{+} \\
    0, & \forall i \in \mathcal{N}_{\text{sub}}
\end{cases}
,\quad \forall t \in \mathcal{T}
\label{eq:connectivity-2}
\end{gather}
Here, $\beta_{ij,t}$ is a binary variable indicating whether bus $i$ serves as the upstream node for bus $j$ at time $t$.

Finally, the switching constraint limits the frequency of switch operations to extend equipment lifespan and maintain operational stability. This is described in \eqref{eq:switch}.
\begin{gather}
\begin{cases}
    |\alpha_{ij,t} - \alpha_{ij,t-1}| \leq \gamma_{ij,t}, & \forall (i,j) \in \mathcal{E},\ \forall t \in \mathcal{T} \\
    \sum_{t \in \mathcal{T}} \gamma_{ij,t} \leq \gamma_{ij}^{\max}, & \forall (i,j) \in \mathcal{E}
\end{cases}
\label{eq:switch}
\end{gather}
In this formulation, the auxiliary binary variable $\gamma_{ij,t}$ indicates whether a switching action occurs on branch $(i,j)$ at time $t$, and $\gamma_{ij}^{\max}$ represents the maximum allowable number of switch operations over the entire time horizon.

\subsection{Two-Stage Stochastic VVO Formulation}
\subsubsection{Compact Formulation}
To enhance clarity, all decision variables and uncertain parameters are categorized into distinct groups and represented as compact vectors. The vector $\bm{\xi}$ comprises the uncertain parameters, while the vector $\bm{y}$ contains scenario-dependent decision variables that capture the system’s recourse behavior. Vectors $\bm{x}$ and $\bm{z}$ represent the primary first-stage decision variables and the auxiliary first-stage variables functionally dependent on $\bm{x}$, respectively; both remain invariant across all scenarios.
\begin{gather}
\left|
\begin{aligned}
\bm{\xi} &: \left\{ \widehat{Pg}_{i,t}^\varphi, \widehat{Pd}_{i,t}^\varphi, \widehat{Qd}_{i,t}^\varphi \right\} \\
\bm{y} &: \left\{ Pg_{i,t}^\varphi, Qg_{i,t}^\varphi, P_{ij,t}^\varphi, Q_{ij,t}^\varphi, U_{i,t}^\varphi, U_{i,t}^{\mathrm{avg}}, \delta_{i,t}^\varphi, z_{i,t}^\varphi \right\} \\
\bm{x} &: \left\{ x_{ij,t}^m, \alpha_{ij,t} \right\} \\
\bm{z} &: \left\{ y_{ij,t}^m, \beta_{ij,t}, \gamma_{ij,t}, \tau_{ij,t}, \rho_{ij,t} \right\}
\end{aligned}
\right.
\label{compact_vector}
\end{gather}

Leveraging the compact vector notation \eqref{compact_vector}, the VVO model defined in \eqref{eq:obj}--\eqref{eq:switch} is reformulated as a two-stage stochastic programming problem. The 2S-VVO model is expressed as follows:
\begin{align}
    \min_{\bm{x},\, \bm{z}} \quad 
    & \bm{c}^\mathsf{T} \bm{z} + \mathbb{E}_{\bm{\xi}} \big[ Q(\bm{x}, \bm{\xi}) \big] \notag \\
    \text{s.t.} \quad 
    & \bm{A} \bm{x} + \bm{B} \bm{z} \leq \bm{d}, \quad 
      \bm{x} \in \mathcal{X}, \quad \bm{z} \in \mathcal{Z}, \label{eq:compact1} \\
    & Q(\bm{x}, \bm{\xi}) := 
    \min_{\bm{y} \in \mathcal{Y}} 
    \big\{ \bm{q}^\mathsf{T} \bm{y} \;\big|\; 
    \bm{W} \bm{y} \geq \bm{h}(\bm{\xi}) - \bm{H} \bm{x} \big\} \label{eq:compact2}
\end{align}

\subsubsection{Scenario-Based 2S-VVO Formulation}
The expectation term $\mathbb{E}_{\bm{\xi}}[Q(\bm{x}, \bm{\xi})]$ is typically intractable to evaluate directly due to the continuous nature of the uncertainty space. To address this challenge, the uncertainty is approximated by a finite number of scenarios, leading to a deterministic reformulation of the 2S-VVO problem, commonly referred to as the extensive form \cite{frau2012stochastic}.
Consider a scenario set $\mathcal{S} = \{\bm{\xi}_{1}, \ldots, \bm{\xi}_{S} \}$, sampled from a given probability distribution $\mathbb{P}$, with each scenario $\bm{\xi}_s$ assigned a probability $\pi_{s}$ for $s \in \mathcal{S}$. The second-stage subproblem is replicated for each scenario, and the expected recourse cost is calculated as a weighted sum over all scenarios. Thus, the extensive form of the 2S-VVO model is written as:
\begin{align}
    \min_{\bm{x},\, \bm{z},\, \{\bm{y}_s\}} \quad 
    & \bm{c}^\mathsf{T} \bm{z} + \sum_{s \in \mathcal{S}} \pi_s\, \bm{q}^\mathsf{T} \bm{y}_s \notag \\
    \text{s.t.} \quad 
    & \bm{A} \bm{x} + \bm{B} \bm{z} \leq \bm{d}, ~~ 
      \bm{x} \in \mathcal{X}, ~~ \bm{z} \in \mathcal{Z}, \label{eq:extensive} \\
    & \bm{W} \bm{y}_s \geq \bm{h}(\bm{\xi}_s) - \bm{H} \bm{x}, ~~ \bm{y}_s \in \mathcal{Y}, ~~ \forall s \in \mathcal{S} \notag
\end{align}
where $\bm{y}_s$ denotes the scenario-specific recourse variables under scenario $\bm{\xi}_s$. 

It is worth noting that the size of the extensive form \eqref{eq:extensive} grows linearly with the number of scenarios in terms of both variables and constraints. Consequently, when a large scenario set is required to adequately capture system uncertainty, the resulting 2S-VVO model can become computationally prohibitive. This challenge is further exacerbated by the recourse problem of VVO, which is formulated as an MILP with binary decision variables in $\bm{y}_{s}$, amplifying the combinatorial complexity. Such scalability issues pose a significant barrier to the practical deployment of the 2S-VVO framework in real-world distribution systems.

\section{Neural Two-Stage Stochastic Optimization}

The 2S-VVO model results in a large-scale MILP formulation, particularly due to the incorporation of complex network reconfiguration constraints and voltage violation duration limits. The associated second-stage resource problem is also formulated as an MILP, which renders conventional decomposition approaches, such as Benders decomposition \cite{oliveira2014accelerating} and column-and-constraint generation \cite{zeng2013ccg}, less effective or impractical. To address the computational challenge, this section introduces a neural-embedded optimization framework designed to efficiently solve the complex 2S-VVO problem.

\subsection{Conceptual Insight}

The second-stage recourse problem defines a mapping from the first-stage decision $\bm{x}$ and scenario realization $\bm{\xi}$ to a scalar objective value, denoted as $Q(\bm{x}, \bm{\xi})$. Given that the recourse problem is an MILP and constitutes the primary source of computational burden, a neural network can be employed to approximate the mapping $Q(\bm{x}, \bm{\xi})$. This approach facilitates efficient evaluation of candidate decisions across a large set of scenarios, thereby eliminating the need to repeatedly solve the recourse problem.

A well-designed surrogate neural network can be seamlessly embedded within the VVO framework, acting as a proxy model that encapsulates the impact of numerous uncertainty scenarios without requiring their explicit enumeration. Since $Q(\bm{x}, \bm{\xi})$ produces a scalar output, the surrogate network can remain compact and computationally efficient, thereby facilitating its integration into the optimization process.

\subsection{Neural Network–Constrained VVO Model}

We replace the explicit second-stage recourse with a \emph{structure-exploiting} surrogate that is constrained by a learned module. The resulting formulation preserves the computational footprint of a first-stage optimization while injecting decision-aware information distilled from the scenario set and the network topology:
\begin{align}
    \min_{\bm{x}, \bm{z}} \quad 
    & \bm{c}^\mathsf{T} \bm{z} + J(\psi) \notag \\
    \text{s.t.} \quad 
    & \bm{A} \bm{x} + \bm{B} \bm{z} \leq \bm{d}, \quad 
      \bm{x} \in \mathcal{X}, \quad \bm{z} \in \mathcal{Z},  \label{nn-vvo-model} \\
    & \psi = \mt{NN}\!\left(\bm{x}, \{\bm{\xi}_{s}\}_{s=1}^{S}\right), \notag
\end{align}
where $J(\psi)$ rescales the surrogate output back to the physical cost level via the affine map:
\begin{equation}
    J(\psi) = \psi \cdot (J_{\max} - J_{\min}) + J_{\min}.
\end{equation}

The surrogate $\mt{NN}(\cdot)$ is \emph{decision-aware}: it takes the first-stage vector $\bm{x}$ and a compressed representation of the uncertainty as inputs, and returns a prediction of the expected second-stage cost. This design allows the optimizer to “see” how $\bm{x}$ interacts with network physics and scenario variability without re-solving the full recourse for each scenario.

To endow the surrogate with the correct inductive biases, we factor $\mt{NN}$ into a topology-aware encoder and a lightweight evaluator in \eqref{eq:nn-overview}:
\begin{equation}
    \mt{NN} \left(\bm{x}, \{\bm{\xi}_{s}\}_{s=1}^{S}\right) = \Phi_{\text{Main}}\left(\bm{x}, \Phi_{\text{Scn}}\left(\{\bm{\xi}_{(s)}\}_{s=1}^{S}\right)\right).
    \label{eq:nn-overview}
\end{equation}

Only $\Phi_{\mt{Main}}$ is embedded within the optimization model and the encoder $\Phi_{\mt{Scn}}$ runs as a pure forward pass. This decoupling lets us deploy an expressive encoder without increasing solver complexity, while keeping the embedded component compact (see Section~\ref{section:nn_embed} for the embedding mechanism). 

\vspace{-1em}
\subsection{Neural Network Design Principles}
\vspace{-0.5em}
Our architecture is designed around three requirements: (i) \emph{topology awareness} ensuring that the surrogate model respects power-flow locality and permutation invariance; (ii) \emph{sample efficiency}, achieved by leveraging structural priors rather than relying solely on large data volumes; and (iii) \emph{embed-ability}, requiring the embedded component in \eqref{nn-vvo-model} to strike a balance between predictive fidelity and tractable solver integration.

\subsubsection{Scenario encoder $\Phi_{\text{Scn}}$}
Distribution networks admit a canonical graph representation (buses as nodes, lines as edges). After extensive design-space exploration, we adopt a graph convolutional network (GCN) for $\Phi_{\text{Scn}}$ because message-passing over the feeder topology provides the right inductive bias to aggregate scenario signals along electrically meaningful paths. This choice (i) enforces permutation invariance over node orderings, (ii) captures locality and hierarchical interactions induced by the network, and (iii) yields a topology-aware uncertainty summary that generalizes across operating points and mild topology perturbations. In practice, the GCN-based encoder markedly improves sample efficiency and generalization ability by extracting informative, physics-aligned features from $\{\bm{\xi}_{s}\}_{s=1}^{S}$ before any decision interaction.

\subsubsection{Main evaluator $\Phi_{\text{Main}}$}
To integrate the surrogate into \eqref{nn-vvo-model} without compromising tractability, we implement $\Phi_{\text{Main}}$ as a compact multilayer perceptron (MLP). This choice reflects a deliberate trade-off: the MLP is sufficiently expressive to model the interaction between the first-stage decision $\bm{x}$ and the encoded uncertainty $\bm{\zeta}$, yet remains amenable to efficient embedding within the optimization (see Section~\ref{section:nn_embed}). Empirically, this yields a favorable accuracy–complexity balance: the topology is learned where it matters (in $\Phi_{\text{Scn}}$), while the embedded component stays lightweight and solver-friendly.

\begin{figure*}[!ht]
    \vspace{-2em}
    \centering
    \includegraphics[width=2\columnwidth]{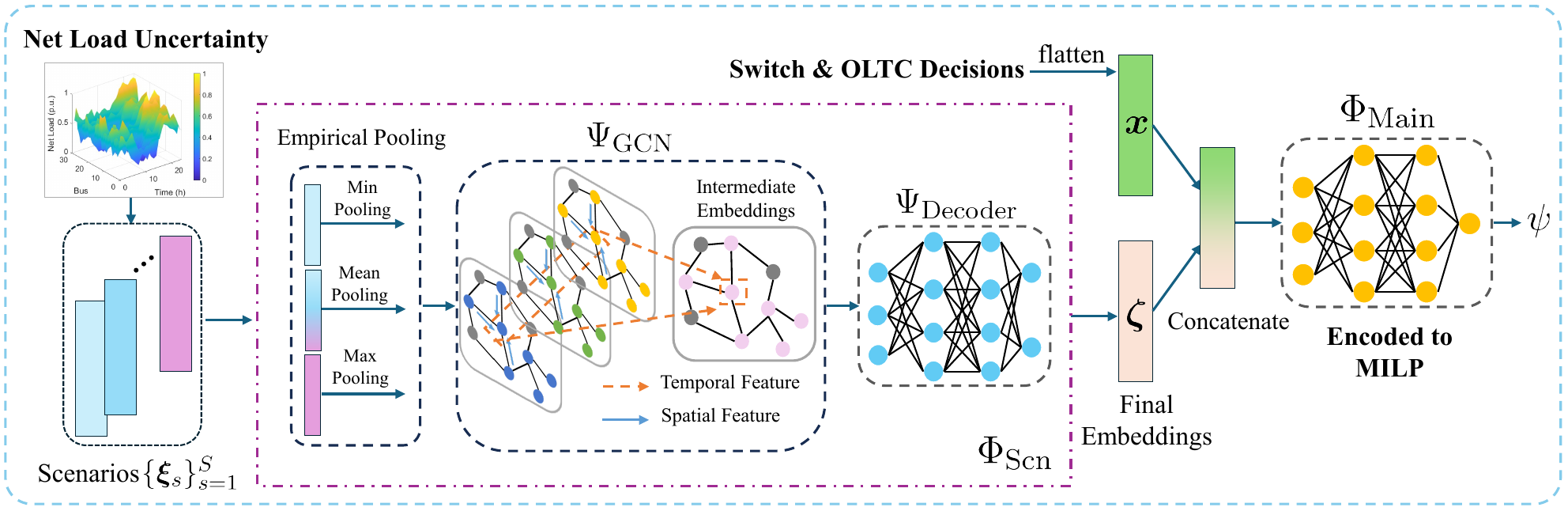}
    \vspace{-0.5em}
    \caption{Architecture diagram of the proposed neural network.}
    \label{fig:nnr-architecture}
    \vspace{-1em}
\end{figure*}

The overall framework is summarized in Fig.~\ref{fig:nnr-architecture}. The encoder $\Phi_{\text{Scn}}$ operates solely as a forward module and may adopt an arbitrarily rich structure, whereas only the compact $\Phi_{\text{Main}}$ enters the optimization, ensuring strong generalization and high sample efficiency without inflating the VVO solve time.

\vspace{-1em}
\subsection{Neural Network Formulation}
\vspace{-0.5em}

\subsubsection{Uncertainty Scenario Preprocessing} 
The uncertainty scenarios, denoted as $\{ \bm{\xi}_{(1)}, \bm{\xi}_{(2)}, \cdots, \bm{\xi}_{(S)} \}$ where each $\bm{\xi}_{(s)} \in \mathbb{R}^{N \times T}$, are first processed through a sequence of empirically-motivated pooling operations to generate a compressed representation suitable for input to the spatiotemporal GCN, thereby enabling its application to an arbitrary number of scenarios. Specifically, three types of pooling operations (i.e., min, mean, and max pooling layers) are employed. This process can be described in \eqref{eq:pooling_layers}:
\begin{align}
    \tilde{\bm{\xi}} = \left[
    \min_{s=1,\dots,S} \bm{\xi}_{(s)},~ 
    \frac{1}{S} \sum_{s=1}^{S} \bm{\xi}_{(s)},~ 
    \max_{s=1,\dots,S} \bm{\xi}_{(s)} 
    \right] \in \mathbb{R}^{N \times T \times F_{\text{in}}} 
    \label{eq:pooling_layers}
\end{align}
Here, $\min$ and $\max$ denote element-wise operations, $\tilde{\bm{\xi}}$ represents the compressed uncertainty feature, $N$ is the number of system nodes, and $T$ is the number of time periods. The entries corresponding to non-uncertainty nodes are set to zero. $F_{\text{in}}$ denotes the resulting feature dimension. 
Subsequently, to highlight the spatiotemporal feature, $\tilde{\bm{\xi}}$ is reshaped as follows:
\begin{align}
    \tilde{\bm{\xi}} \in \mathbb{R}^{N \times T \times F_{\text{in}}}  \xrightarrow{\text{reshape}} \tilde{\bm{\xi}}^{(1)}, \cdots, \tilde{\bm{\xi}}^{(T)} \in \mathbb{R}^{N \times F_{\text{in}}}
\end{align}

\subsubsection{Spatiotemporal GCN for Scenario Embedding}
First, the spatial dependencies of uncertainties are captured. At each time step, a GCN operation $\Psi_{\text{GCN}}$ is applied independently. Specifically, at time $t$, we compute the spatial embedding as:
\begin{equation}
\mathbf{H}^{(t)} = \sigma\left(\tilde{\mathbf{D}}^{-1/2} \tilde{\mathbf{A}}\tilde{\mathbf{D}}^{-1/2}\tilde{\bm{\xi}}^{(t)} \mathbf{W}\right),
\end{equation}
where $\mathbf{H}^{(t)} \in \mathbb{R}^{N \times d_{\text{in}}}$ is the spatial embedding at time step $t$, $\tilde{\mathbf{A}} = \mathbf{A} + \mathbf{I}$ denotes the adjacency matrix with added self-loops, $\mathbf{A}$ is the adjacency matrix of the distribution network with all switches connected, and $\mathbf{I}$ is an identity matrix.
$\tilde{\mathbf{D}}$ is the diagonal degree matrix with entries $\tilde{\mathbf{D}}_{ii} = \sum_j \tilde{\mathbf{A}}_{ij}$, $\mathbf{W} \in \mathbb{R}^{F_{\text{in}} \times d_{\text{in}}}$ is a learnable weight matrix, and $\sigma(\cdot)$ is a nonlinear activation function.

Next, we consider temporal dynamics by stacking spatial embeddings for each node. For node $n$, let $\mathbf{Z}_0^{(n)} := [\mathbf{H}_n^{(1)}; \mathbf{H}_n^{(2)}; \dots; \mathbf{H}_n^{(T)}] \in \mathbb{R}^{T \times d_{\text{in}}}$ denote the temporal sequence of spatial embeddings.

A stack of $L$ temporal convolutional layers (i.e., the 1D-CNN) processes these sequences independently for each node. Each temporal convolutional layer $\ell = 1, \dots, L$ operates as follows:
\begin{equation}
\mathbf{Z}_\ell^{(n)} = \sigma\left(\mathbf{Z}_{\ell-1}^{(n)} * \mathbf{W}_\ell + \mathbf{b}_\ell\right),
\end{equation}
where $\mathbf{Z}_{\ell-1}^{(n)} \in \mathbb{R}^{T_{\ell-1} \times d_{\ell-1}}$ is the input tensor to layer $\ell$, $\mathbf{W}_\ell \in \mathbb{R}^{k_\ell \times d_{\ell-1} \times d_\ell}$ is the 1D convolutional kernel with kernel size $k_\ell$ and $d_\ell$ output channels, $\mathbf{b}_\ell \in \mathbb{R}^{d_\ell}$ is the bias term, and $*$ denotes the convolution operation along the temporal dimension.

After applying $L$ convolutional layers, the resulting tensor is aggregated into a fixed-size node embedding by concatenating the node features as follows:
\begin{equation}
    \widetilde{\mathbf{Z}} = [\mathbf{Z}_L^{(1)}; \mathbf{Z}_L^{(2)}; \dots; \mathbf{Z}_L^{(N)}] \in \mathbb{R}^{1 \times N d_{\text{out}}}.
\end{equation}

Subsequently, the embedding $\widetilde{\mathbf{Z}}$ is fed into a fully-connected MLP, referred to as the $\Psi_{\mt{Decoder}}$, to perform dimensionality reduction. This process yields the final output representation $\bm{\zeta}$ as: $\bm{\zeta} = \Psi_{\mt{Decoder}}(\widetilde{\mathbf{Z}})$.

In summary, the scenario embedding network $\Phi_{\mt{Scn}}$ is represented as:
\vspace{-0.75em}
\begin{equation}
    \bm{\zeta} = \Phi_{\mt{Scn}}(\{\bm{\xi}_{s}\}_{s=1}^{S}) = \Psi_{\mt{Decoder}}
    \left(
        \Psi_{\mt{GCN}}(\{\bm{\xi}_{s}\}_{s=1}^{S})
    \right)
\end{equation}
\vspace{-1.5em}

\subsubsection{Main Network for Optimization Integration}

The main network $\Phi_{\text{Main}}$, structured as an MLP with Rectified Linear Unit (ReLU) activation function, estimates the expected second-stage cost based on the scenario embedding $\bm{\zeta}$ and decision vector $\bm{x}$. Formally, it computes: $\psi = \Phi_{\text{Main}}(\bm{\zeta}, \bm{x})$.

Once the uncertainty embedding $\bm{\zeta}$ is fixed, the MLP can be encoded into an MILP formulation using standard ReLU-to-linear constraint transformations. This allows direct optimization over the decision vector $\bm{x}$ while enforcing operational constraints such as switch and OLTC limits, into the optimization problem. By integrating the neural network with these constraints, we establish a surrogate optimization framework suitable for efficient decision-making.

\vspace{-1.25em}
\subsection{Encoding Neural Networks into Optimization Framework} 
\vspace{-0.25em}
\label{section:nn_embed}

We consider an $\ell$-layer fully connected feedforward neural network that maps an input vector $\bm{x}$ to a scalar output $\psi$. The network architecture is defined through the following forward propagation equations:
\begin{align}
    &\mathbf{h}^{1} = \sigma \left( \mathbf{W}^{1} \bm{x} + \mathbf{b}^{1} \right), \\
    &\mathbf{h}^{m} = \sigma \left( \mathbf{W}^{m} \mathbf{h}^{m-1} + \mathbf{b}^{m} \right), \quad m = 2, \ldots, \ell - 1, \\
    &\psi = \mathbf{W}^{\ell} \mathbf{h}^{\ell - 1} + \mathbf{b}^{\ell},
\end{align}
where $\mathbf{h}^{m}$ represents the activation vector of layer $m$, $\mathbf{W}^{m}$ and $\mathbf{b}^{m}$ denote the weight matrix and bias vector for layer $m$, respectively, and $\sigma(\cdot)$ is a nonlinear activation function. 

When employing the ReLU activation function, the activation of the $i$-th neuron in layer $m$ can be expressed as:
\begin{equation}
    h_{i}^{m} = \text{ReLU} \left( \sum_{j=1}^{N_{j}^{m}} w_{ij}^{m} h_{j}^{m-1} + b_{i}^{m} \right),
    \label{eq-relu}
\end{equation}
where $w_{ij}^{m}$ represents the weight connecting neuron $j$ in layer $(m-1)$ to neuron $i$ in layer $m$, $b_{i}^{m}$ is the corresponding bias term, and $N_{j}^{m}$ denotes the number of neurons in the preceding layer.

To incorporate the neural network into an optimization framework, we reformulate the ReLU activation function as a set of MILP constraints. This reformulation introduces auxiliary variables $\hat{h}_{i}^{m}$ and $\check{h}_{i}^{m}$ to represent the positive and negative components of the pre-activation value, respectively, along with a binary indicator variable $\mu_i^m$ that determines whether neuron $i$ in layer $m$ is activated:
\begin{subequations}
\begin{align}
    &\sum_{j=1}^{N_{j}^{m}} w_{ij}^{m} h_{j}^{m-1} + b_i^{m} = \hat{h}_i^m - \check{h}_i^m, \\
    &0 \leq \hat{h}_i^m \leq \mu_i^m M, \\
    &0 \leq \check{h}_i^m \leq (1 - \mu_i^m) M, \\
    &\mu_i^m \in \{0, 1\},
\end{align}
\label{eq-relu-milp}
\end{subequations}
\noindent where $M$ is a sufficiently large constant. To improve numerical stability and computational efficiency, we can replace the uniform big-$M$ constants with neuron-specific bounds that are derived from the bounded input domain and the hierarchical structure of the network. This approach yields tighter relaxations and mitigates numerical issues commonly associated with large uniform constants.

\vspace{-1em}
\subsection{Data Generation Strategy}
To construct the training dataset, we first define a set of scenarios that capture the stochastic nature of load demand and DER generation. Each stochastic programming-based VVO problem is referred to as an \emph{instance}, where one instance consists of multiple scenarios. Each sampling  instance is composed of 10 scenarios. Solving one such instance yields a training sample for the proposed neural network. By repeating this process, we generate a total of 5,000 samples for model training. For testing, new instances are generated with varying numbers of scenarios, in order to evaluate both the effectiveness and generalization capability of the proposed model.

The scenario generation procedure is as follows: first, real-world load shapes and PV generation profiles are collected from reference \cite{oedi_5773}. Then, random uniform perturbations are applied to the load and DER allocation across network nodes, ensuring stochastic variability among nodes. On average, solving a single instance requires approximately 35 seconds. With four parallel samplers, generating 5,000 samples takes about 12.16 hours in total.

\vspace{-1em}
\subsection{Accelerated Sampling Strategy}
The efficiency of sample generation is further enhanced by reformulating the 2S-VVO into a sequence of multi-period OPFs, thereby reducing the number of binary variables and significantly accelerating computation. In this framework, the first-stage binary decisions are randomly sampled and enforced to satisfy the operational feasibility conditions through an auxiliary problem, denoted as formulation \eqref{aux-problem}. Owing to its simplicity, Problem \eqref{aux-problem} can be solved with almost negligible computational cost. Once the first-stage decisions are determined, the corresponding recourse problems are solved scenario-by-scenario, yielding approximate yet high-quality solutions to stochastic VVO instances. This procedure substantially improves the efficiency of data sampling, as evidenced in Table~\ref{tab:acceleration}.

\vspace{-2em}
\begin{subequations}
    \begin{align}
        \min_{\bm{x},\, \bm{z}} \quad 
        & \bm{r}^\mathsf{T} \bm{x} \\
        \text{s.t.} \quad & \|\bm{x} - \bm{x}^{*}\|_1 \leq \eta N T, \\
        & \bm{A} \bm{x} + \bm{B} \bm{z} \leq \bm{d}, \\
        & \bm{x} \in \mathcal{X},\; \bm{z} \in \mathcal{Z},
    \end{align}
    \label{aux-problem}
\end{subequations}
\vspace{-1.75em}

\noindent where $\bm{x}^{*}$ denotes the kernel VVO solution and $\bm{r}$ is a random vector uniformly sampled from $[-1,1]$ to introduce stochastic variation. The parameter $\eta \in (0,1)$ defines the perturbation radius, with $\eta = 0.2$ commonly adopted to restrict deviations to 20\% of the discrete variables. We define \emph{Kernel VVO} solutions as representative VVO instances obtained by solving deterministic VVO problems under selected scenarios, which serve as proxies for capturing diverse solution patterns. 

\input{table/table-sampling-time}

\vspace{-1em}
\section{Numerical Study}\label{sec:case_study}
\vspace{-0.25em}
\subsection{Numerical Study Setup}
\begin{figure}[t]
    \vspace{-0.5em}
    \centering
    \includegraphics[width=0.9\columnwidth]{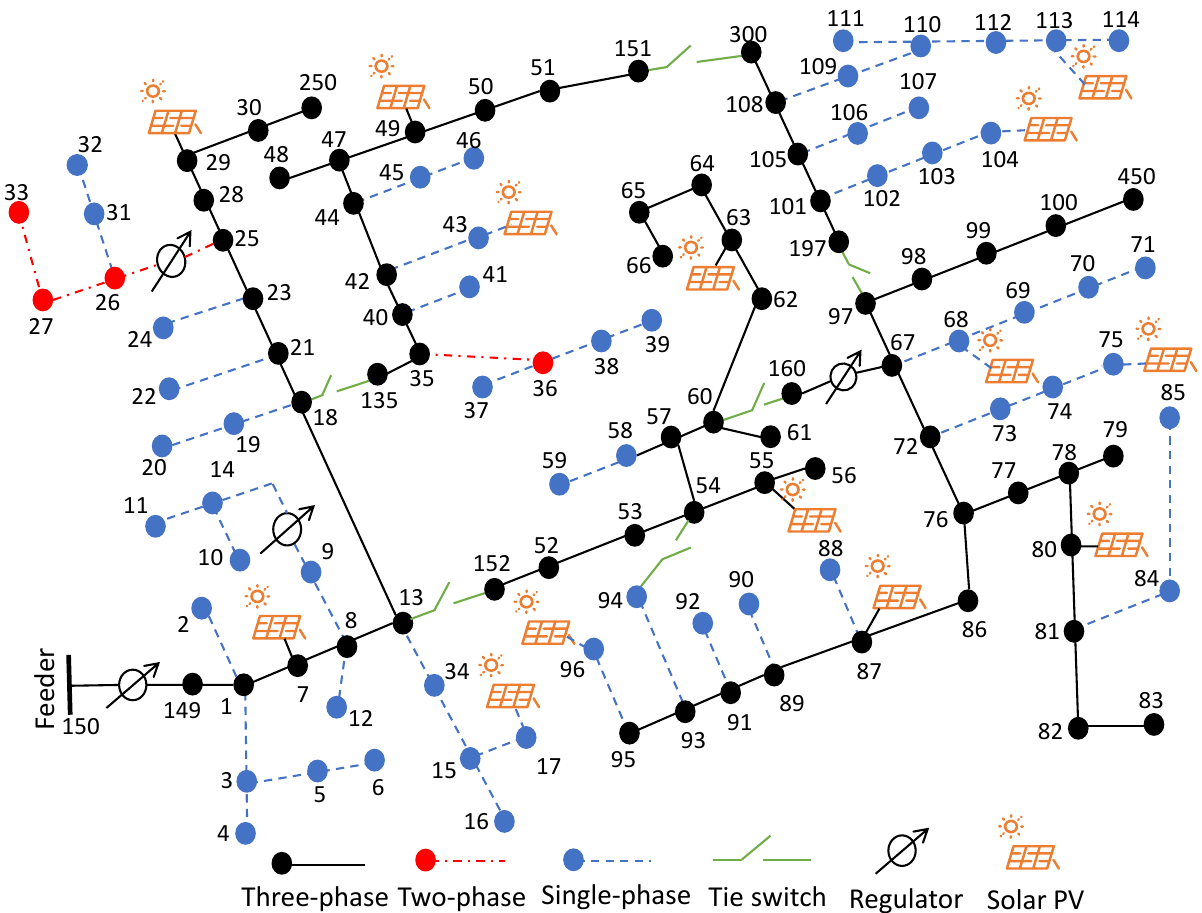}
    \vspace{-0.5em}
    \caption{Diagram of the modified IEEE 123-bus system.}
    \label{fig:123bus}
    \vspace{-1.5em}
\end{figure}

\vspace{-0.5em}
The proposed method is validated on the modified IEEE 123-bus unbalanced distribution system, as illustrated in Fig.~\ref{fig:123bus}. The system comprises four voltage regulators, six tie switches, and fourteen solar PV systems. The hyperparameters used in the case studies are summarized in Table~\ref{tab:sys_param}. System uncertainties arise from three-phase load and solar PV generation. Uncertainty scenarios are derived from the open-source dataset in \cite{oedi_5773} developed by the National Renewable Energy Laboratory (NREL), which contains multi-year historical load and PV generation curves. The stochastic programming scenarios are generated by randomly sampling from this dataset. For the MILP solver, Gurobi~12.0.1 is employed with a three-hour time limit and a MIP gap tolerance of 0.01\%.
\input{table/table-parameters}

\input{table/table-nn-parameters}

The neural network surrogate is trained in PyTorch, and the resulting network-based optimization problems are formulated using the OMLT toolbox~\cite{ceccon2022omlt}. OMLT integrates big-M acceleration strategies, enabling substantial computational speedups. The neural network hyperparameters are summarized in Table~\ref{tab:nn-parameters}. All simulations are conducted on a workstation equipped with an Intel i9-9900X @ 3.5\,GHz processor and 64\,GB RAM.

Two benchmark methods are considered for comparison:
\begin{enumerate}
    \item[(1)] \textbf{Gurobi}: Directly solving the 2S-VVO problem using the extensive MILP formulation.
    \item[(2)] \textbf{PH}: The progressive hedging method \cite{PH-2}, a state-of-the-art decomposition approach for addressing large-scale stochastic programming problems.
\end{enumerate}

Solution quality is measured by the optimality gap (hereafter referred to as the gap), computed relative to the exact Gurobi solution. All metrics are evaluated over 100 test instances.

\vspace{-1.25em}
\subsection{Computation Time and Solution Quality}
\vspace{-0.25em}

\input{table/table-comparison}

The performance of the proposed method is evaluated against \textbf{Gurobi} and \textbf{PH}. Table~\ref{tab-comparison} summarizes the results, where \textit{\#Scen} denotes the number of uncertainty scenarios in a 2S-VVO instance. Tests are conducted under varying levels of monitoring coverage, denoted as \textbf{LM}, \textbf{MM}, and \textbf{FM}, whose definitions are provided in Table~\ref{tab:def}. Increasing the number of monitoring nodes or scenarios increases the problem size by adding constraints and binary variables, thereby increasing computational complexity.

\input{table/table-def}

Across all model scales and scenario sizes, the proposed method yields objective values closely matching those of Gurobi, maintaining an optimality gap below $0.27\%$. The overall average gap is $0.216\%$, with the lowest value of $0.180\%$ for the 123\_Bus\_LM case with 10 scenarios and the highest value of $0.262\%$ for the 123\_Bus\_FM case with 100 scenarios. Although larger models and more scenarios tend to slightly increase the gap, the relationship is not strictly monotonic. For instance, the 123\_Bus\_MM case with 10 scenarios has a gap of $0.193\%$, slightly higher than the $0.186\%$ for the 123\_Bus\_FM case with the same number of scenarios.

In terms of computational efficiency, the proposed approach achieves significant time savings. Average solution times are 1720.0s for Gurobi, 265.4s for PH, and 32.1s for the proposed method. This corresponds to a $50.1\times$ speedup over Gurobi. Notably, the runtime remains nearly constant at about 32s regardless of model size or scenario count, since the size of surrogate-based MILP is scenario-independent. This scalability makes the method particularly effective for large-scale problems: the speedup is modest $1.1\times$ for the 123\_Bus\_LM case with 10 scenarios but grows dramatically to $200.6\times$ for the 123\_Bus\_FM case with 100 scenarios.

\vspace{-1.25em}
\subsection{Computation Efficiency and Optimality Trade-Off}
\vspace{-0.25em}
While the proposed method achieves substantial improvements in computational efficiency, it yields slightly larger optimality gaps compared to the Gurobi solver. To further quantify the efficiency gains and examine the trade-off between solution quality and computational time, two supplementary experiments are conducted. First, we measure the time required for Gurobi to reach the same optimality gap as that produced by the proposed method; this test is referred to as \textbf{Gurobi-to-NN-gap}. Second, we use the solution obtained from the proposed method as a warm-start for Gurobi and evaluate the time required for Gurobi to refine this solution to full optimality; this test is referred to as \textbf{NN-to-optimal}, where \textbf{NN} denotes the proposed neural network-based approach. These analyses provide a direct measure of the computational efficiency improvements when targeting a fixed solution quality.

\begin{figure}[tbp]
    \centering
    \vspace{-0.75em}
    \includegraphics[width=0.7\columnwidth]{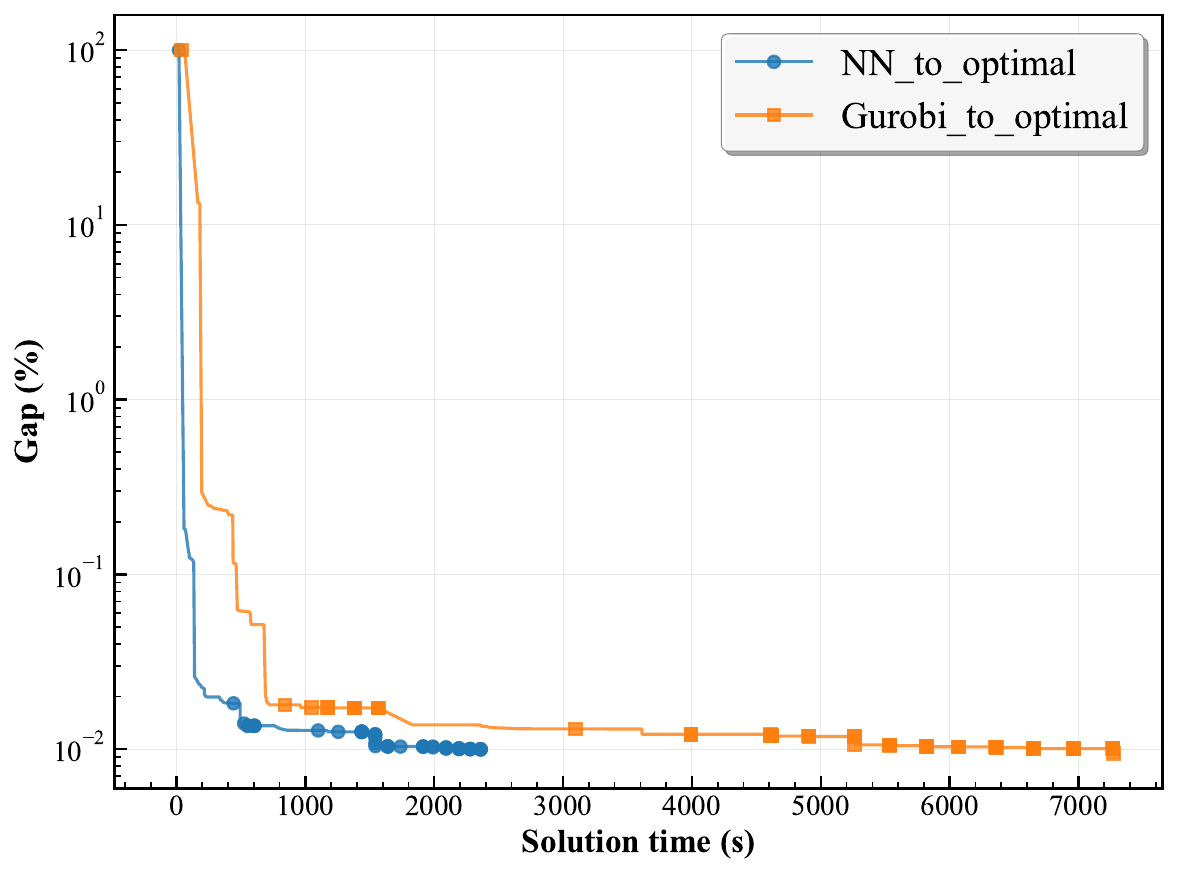}
    \vspace{-1em}
    \caption{Evolution of the optimality gap over computation time for the 123\_Bus\_FM case with 100 scenarios.}
    \vspace{-1.75em}
    \label{fig-time-gap}
\end{figure}

The 123\_Bus\_FM case under 10, 50, and 100 scenarios is used for testing. The results are summarized in Table~\ref{tab:new-performance}, and Fig.~\ref{fig-time-gap} depicts the evolution of the optimality gap over time for the Gurobi-to-optimal and NN-to-optimal cases under the 100-scenario setting.

First, the comparison between the proposed NN approach and the Gurobi-to-NN-gap test highlights the time required to achieve equivalent solution quality. With 100 scenarios, the NN method obtains a high-quality solution in just 37.6s, whereas Gurobi requires 2128.3s to reach a solution of comparable quality, representing a $56.6\times$ reduction in computation time. A similar trend is observed for 50 scenarios, where the NN method requires 34.3s compared to 657.4s for Gurobi. For the 10-scenario setting, the advantage is somewhat reduced, with NN taking about 28.6s compared to 102.4s for Gurobi.

Second, comparing NN-to-optimal with Gurobi-to-optimal quantifies the benefit of using the proposed method’s solution as a warm start. In the 100-scenario case, refining the NN solution to a MIP gap of 0.01\% takes 2367.2s, whereas Gurobi without warm start takes 7289.5s. This corresponds to a reduction of nearly 70\% in computation time. At 50 scenarios, refining the NN solution to optimal requires 932.1s, compared to 3126.3s for Gurobi, while at 10 scenarios the times are 95.6s and 498.6s, respectively. This corresponds to an improvement of approximately 80\%. Although the absolute time savings are smaller in lower-dimensional settings, the relative efficiency gains remain substantial across all scales.

\input{table/table-refine-solution}

\vspace{-3mm}
\subsection{Scalability Analysis of the Proposed Method}

\input{table/table-scale}

The scalability of the proposed method is examined for the 123\_Bus\_MM configuration with scenario counts of 100, 200, 500, and 1000. Results are given in Table~\ref{table:table-scale}.

The Gurobi solver, serving as the baseline, exhibits a significant increase in computation time as the number of scenarios grows, that is from 2,616 seconds at 100 scenarios to 26,819 seconds at 1,000 scenarios. This trend reflects the linear growth of the extensive-form problem with scenario count.  

The PH method offers a considerable reduction in computation time compared to Gurobi, with runtimes ranging from 379 to 2,520 seconds. Importantly, the solution quality remains high, as the optimality \emph{gap} consistently stays below $0.20\%$. The resulting speedup over Gurobi lies between \textbf{6.91$\times$} and \textbf{10.65$\times$}, indicating significantly improved efficiency and better scalability relative to the extensive-form baseline.  

By contrast, the proposed neural network-based optimization method demonstrates highly favorable scalability. Across all tested scenarios, solution times remain tightly bounded between 34.5 and 51.1 seconds, demonstrating only minimal sensitivity to scenario count. This is attributed to the fixed dimensionality of the surrogate MILP formulation, wherein scenario variations influence only the neural network inputs rather than the optimization model’s size. In terms of solution quality, the proposed method yields \emph{gap} values between $0.255\%$ and $0.295\%$, slightly higher than those of PH but still well within practical tolerances. The computational gains, however, are substantial: speedups range from \textbf{75.8$\times$} to \textbf{524.8$\times$}, increasing sublinearly with the number of scenarios, demonstrating the method’s suitability for large-scale stochastic VVO in unbalanced distribution networks.

\vspace{-1em}
\section{Conclusion} 
\vspace{-0.5em}
We propose a neural two-stage stochastic optimization framework for solving VVO in three-phase unbalanced distribution systems with network reconfiguration. By embedding a deep neural network surrogate of the second-stage recourse model into the first-stage MILP, the method decouples computational complexity from scenario count, enabling scalable optimization. The formulation also incorporates a voltage violation duration constraint to enhance reliability under uncertain DER outputs. Comparative studies show that the method achieves over 50$\times$ speedup compared to conventional approaches, with negligible optimality loss. These results highlight the effectiveness of the neural-enhanced framework in addressing real-time and large-scale operational challenges in modern power distribution systems.
\vspace{-1em}

\bibliographystyle{IEEEtran}
\bibliography{ref.bib}

\end{document}

%% file: table/table-sampling-time.tex
\begin{table}[t]
    \centering
    \vspace{-1em}
    \caption{Comparison of Training Data Sampling Time}
    \vspace{-0.5em}
    \label{tab:acceleration}
    \begin{tabular}{lccc}
        \toprule
        \textbf{Strategy} & \textbf{Auxiliary Process} & \textbf{Main Process} & \textbf{Total Time} \\
        \midrule
        Non-accelerated  & ---        & 12.16 h & 12.16 h \\
        Accelerated      & 0.46 h  & 4.38 h  & 4.84 h \\
        \bottomrule
    \end{tabular}
    \vspace{-2.25em}
\end{table}

%% file: table/table-parameters.tex
\begin{table}[t]
    \vspace{-1em}
    \caption{Hyperparameters Parameters}
    \vspace{-0.5em}
    \centering
    \renewcommand{\arraystretch}{1.5}
    \begin{tabularx}{\linewidth}{l|X|l} \hline
      Parameter & Description  & Value 
      \\ \hline
      $d_{1}$, $d_{2}$ & Maximum number of voltage violations over total and consecutive time steps. & 8, 4
      \\ \hline
      $M_{i,k}$ & Number of available tap settings. & 5
      \\ \hline
      $Sg_{i}^{\varphi}$, $S_{ik}^{\varphi}$ 
      & Capacity of DERs and distribution lines (kVA). 
      & 50, 2000
      \\ \hline
      $U_{i}^{\min}$, $U_{i}^{\max}$ & Voltage limits of Range B (p.u.). & [0.917, 1.058] 
      \\ \hline
      $\underline{U}_i$, $\overline{U}_i$ & Voltage limits of Range A (p.u.). & [0.95, 1.05]
      \\ \hline
      $\eta$ & Perturbation level in data generation. & 0.2
      \\ \hline
      $\omega^{\mt{der}}$ 
      & Cost coefficient for DER curtailments (\$/kWh).
      & 120
      \\ \hline
      $\omega^{\mt{sw}}$, $\omega^{\mt{oc}}$ 
      & Cost coefficients for switches and OLTCs (\$/switching).
      & 100, 300
      \\ \hline
      $\gamma_{i,k}^{\max}$, $\rho_{i,k}^{\max}$ & Maximum number of operations for switches and OLTCs. & 8, 4
      \\ \hline
    \end{tabularx}
    \vspace{-1em}
    \label{tab:sys_param}
\end{table}

%% file: table/table-nn-parameters.tex
\begin{table}[tbp]
\centering
\caption{Hyperparameters for Neural Networks}
\vspace{-0.5em}
\label{tab:nn-parameters}
\begin{tabularx}{\linewidth}{lX}
\toprule
\textbf{Parameter} & \textbf{Network Settings} \\
\midrule
Batch size & 64 \\
Learning rate & $10^{-3}$ \\
L1 weight penalty & $10^{-4}$ \\
L2 weight penalty & $10^{-5}$ \\
Optimizer & \{Adam, Adagrad\} \\
Dropout & 0.01 \\
\# Epochs & 2000 \\
\midrule
\multicolumn{2}{l}{\textbf{Encoder (Spatio-Temporal GCN)}} \\
\midrule
Number of GCN layers & 2 \\
Spatial aggregation method & Mean \\
Temporal window length & 4 \\
GCN hidden dimension & \{128, 64\} \\
Activation function & ReLU \\
Normalization & LayerNorm \\
Temporal encoding & Positional (learnable embedding) \\
\midrule
\multicolumn{2}{l}{\textbf{Decoder (MLP)}} \\
\midrule
Decoder hidden dimension & \{64, 32\} \\
\midrule
\multicolumn{2}{l}{\textbf{Main Network (MLP embedded into optimization)}} \\
\midrule
ReLU hidden dimension & \{128, 64, 32\} \\
\bottomrule
\end{tabularx}
\vspace{-2em}
\end{table}

%% file: table/table-comparison.tex

\begin{table*}[!t]
\vspace{-3mm}
\renewcommand{\arraystretch}{1.2}
\caption{Performance Comparison of Different VVO Solution Methods in 123-Bus Three-Phase Distribution System}
\label{tab-comparison}
\begin{tabularx}{\textwidth}{llYYYYYYYYYY}
\toprule
\multirow{2}{*}{\vspace{-2mm}\textbf{Cases}} & \multirow{2}{*}{\vspace{-2mm}\textbf{\#Scen}} 
& \multicolumn{3}{c}{\textbf{Avg. Objective (\$)}} 
& \multicolumn{2}{c}{\textbf{Avg. Gap (\%)}} 
& \multicolumn{3}{c}{\textbf{Solution Time (s)}} 
& \multicolumn{2}{c}{\textbf{Speed Up}} \\
\cmidrule(lr){3-5} \cmidrule(lr){6-7} \cmidrule(lr){8-10} \cmidrule(lr){11-12}
& & \textbf{Gurobi} & \textbf{PH} & \textbf{Proposed} 
& \textbf{PH} & \textbf{Proposed} 
& \textbf{Gurobi} & \textbf{PH} & \textbf{Proposed} 
& \textbf{PH} & \textbf{Proposed} \\
\midrule
\multirow{3}{*}{123\_Bus\_LM} 
& $S$ = 10  & 4127.6 & 4131.2 & 4135.1 & 0.086 & 0.180 & 32.1 & 41.3 & \textbf{28.5} & 0.8$\times$ & \textbf{1.1}$\times$ \\
& $S$ = 50  & 4103.2 & 4108.9 & 4111.4 & 0.139 & 0.199 & 213.2 & 92.4 & \textbf{30.9} & 2.3$\times$ & \textbf{6.9}$\times$ \\
& $S$ = 100 & 4096.5 & 4102.8 & 4105.4 & 0.154 & 0.217 & 512.3 & 148.7 & \textbf{31.7} & 3.4$\times$ & \textbf{16.2}$\times$ \\
\midrule
\multirow{3}{*}{123\_Bus\_MM} 
& $S$ = 10  & 4189.1 & 4193.2 & 4197.2 & 0.098 & 0.193 & 208.6 & 52.1 & \textbf{31.6} & 4.0$\times$ & \textbf{6.6}$\times$ \\
& $S$ = 50  & 4164.2 & 4170.2 & 4173.5 & 0.144 & 0.224 & 1059.3 & 189.5 & \textbf{33.5} & 5.6$\times$ & \textbf{31.6}$\times$ \\
& $S$ = 100 & 4151.7 & 4158.6 & 4162.3 & 0.166 & 0.255 & 2616.5 & 378.6 & \textbf{34.5} & 6.9$\times$ & \textbf{75.8}$\times$ \\
\midrule
\multirow{3}{*}{123\_Bus\_FM} 
& $S$ = 10  & 4276.9 & 4280.8 & 4284.8 & 0.091 & 0.186 & 512.5 & 83.6 & \textbf{29.5} & 6.1$\times$ & \textbf{17.4}$\times$ \\
& $S$ = 50  & 4242.3 & 4248.5 & 4252.1 & 0.147 & 0.231 & 3063.1 & 421.5 & \textbf{32.5} & 7.3$\times$ & \textbf{94.2}$\times$ \\
& $S$ = 100 & 4231.6 & 4238.4 & 4242.7 & 0.161 & 0.262 & 7262.6 & 981.2 & \textbf{36.2} & 7.4$\times$ & \textbf{200.6}$\times$ \\
\midrule
\textbf{Average} & -- & 4175.9 & 4181.4 & 4186.0 & 0.132 & \textbf{0.216} & 1720.0 & 265.4 & \textbf{32.1} & 4.9$\times$ & \textbf{50.1}$\times$ \\
\bottomrule
\end{tabularx}
\vspace{-3mm}
\end{table*}

%% file: table/table-def.tex
\begin{table}[hb]
    \vspace{-5mm}
    \centering
    \caption{Definition of Monitoring Levels}
    \vspace{-1mm}
    \renewcommand{\arraystretch}{1.2}
    \begin{tabular}{c|c|c} 
        \hline
        \textbf{Mark} & \textbf{Description} & \textbf{Monitored Buses} 
        \\ \hline
        LM & Lower-level monitoring & \{85,114\}
        \\ \hline
        MM & Medium-level monitoring & \thead{ \{39,66,75,83,85,94,96,\\114,149,160,300,450\} }
        \\ \hline
        FM & Full monitoring & All buses except the substation
        \\ \hline 
    \end{tabular}
    \label{tab:def}
\end{table}

%% file: table/table-refine-solution.tex
\begin{table}[t]
\centering
\vspace{-1em}
\caption{Solution Time for Different Setings on the 123\_Bus\_FM Case}
\vspace{-0.5em}
\label{tab:new-performance}
\renewcommand{\arraystretch}{1.2}
\begin{tabularx}{0.95\linewidth}{l>{\centering\arraybackslash}X>{\centering\arraybackslash}X>{\centering\arraybackslash}X}
\toprule
\multirow{2}{*}{\vspace{-2mm}\textbf{Solution Time (s)$^{\dagger}$}} & \multicolumn{3}{c}{\textbf{Number of Scenarios}} \\
\cmidrule(lr){2-4}
& 10 & 50 & 100 \\
\midrule
Gurobi-to-NN-gap      & 102.4 & 657.4 & 2128.3 \\
NN method                    & 28.6 & 34.3 & 37.6 \\
\midrule
Gurobi-to-optimal      & 498.6 & 3126.3 & 7289.5 \\
NN-to-optimal         & 95.6 & 932.1 & 2367.2 \\
\bottomrule
\end{tabularx}
\\[0.5em]
\raggedright \textit{~~~Notes:} \textit{$^{\dagger}$The MIP Gap for the Gurobi solver is set to 0.01\%} 
\vspace{-0.5em}
\end{table}

%% file: table/table-scale.tex
\begin{table}[t]
\centering
\vspace{-2mm}
\caption{Scalability Performance under Different Scenario Sizes}
\vspace{-1mm}
\label{table:table-scale}
\renewcommand{\arraystretch}{1.2}
\resizebox{\linewidth}{!}{%
\begin{tabularx}{1.02\linewidth}{llllll}
\toprule
\multirow{2}{*}{\vspace{-2mm}\textbf{Method}} & \multirow{2}{*}{\vspace{-2mm}\textbf{Metric}} & \multicolumn{4}{c}{\textbf{Number of Scenarios}} \\
\cmidrule(ll){3-6}
& & 100 & 200 & 500 & 1000 \\
\midrule
\multirow{2}{*}{Gurobi}    
& Objective (\$)   & 4{,}151.7 & 4{,}151.10 & 4{,}149.6 & 4{,}147.9 \\
& Time (s)         & 2{,}616 & 5{,}204 & 13{,}606 & 26{,}819 \\
\midrule
\multirow{3}{*}{PH}       
& Gap (\%)         & 0.166 & 0.160 & 0.172 & 0.181 \\
& Time (s)         & 379 & 680 & 1{,}420 & 2{,}520 \\
& Speedup          & \textbf{6.91$\times$} & \textbf{7.65$\times$} & \textbf{9.58$\times$} & \textbf{10.65$\times$} \\
\midrule
\multirow{3}{*}{Proposed} 
& Gap (\%)         & 0.255 & 0.265 & 0.283 & 0.295 \\
& Time (s)         & 34.5 & 42.2 & 47.5 & 51.1 \\
& Speedup          & \textbf{75.8$\times$} & \textbf{123.3$\times$} & \textbf{286.4$\times$} & \textbf{524.8$\times$} \\
\bottomrule
\end{tabularx}%
}
\vspace{-5mm}
\end{table}